\pdfoutput=1
\documentclass[twocolumn,aps,prb,showpacs,superscriptaddress]{revtex4}
\usepackage{graphicx}
\usepackage{color}
\usepackage{mathtools}
\usepackage{SIunits}

\begin{document}

\title{Unconventional RF Photo-Response from a Superconducting Spiral  Resonator }

\author{Alexander P. Zhuravel}
\affiliation{B. Verkin Institute for Low Temperature Physics and
Engineering, National Academy of Sciences of Ukraine, 61103 Kharkov,
Ukraine}

\author{Cihan Kurter}
\affiliation{Center for Nanophysics and Advanced Materials,
Department of Physics, University of Maryland, College Park,
Maryland 20742-4111 USA}

\author{Alexey V. Ustinov}
\affiliation{Physikalisches Institut and DFG-Center for Functional
Nanostructures (CFN), Karlsruhe Institute of Technology, D-76128
Karlsruhe, Germany}

\author{Steven M. Anlage}
\affiliation{Center for Nanophysics and Advanced Materials,
Department of Physics, University of Maryland, College Park,
Maryland 20742-4111 USA} \affiliation{Physikalisches Institut and
DFG-Center for Functional Nanostructures (CFN), Karlsruhe Institute
of Technology, D-76128 Karlsruhe, Germany}

\date{\today}

\begin{abstract}

Superconducting thin film resonators employing strip geometries show
great promise in $\it RF$/microwave applications due to their
low-loss and compact nature. However, their functionality is limited
by non-linear effects at elevated $\it RF$/microwave powers. Here,
we show that by using a planar spiral geometry carrying parallel
currents in adjacent turns, this limitation can be minimized. We
investigate the $\it RF$ current distributions in spiral resonators
implemented with Nb thin films via Laser Scanning Microscopy. The
$\it RF$ current density profile along the width of the individual
turns of the resonators reveals an unconventional trend: maximum
current in the middle of the structure and decaying towards its
edges. This unusual behavior is associated with the circular nature
of the geometry and the cancellation of magnetic field between the
turns, which is favorable for handling high powers since it allows
the linear characteristics to persist at high $\it RF$ current
densities.
\end{abstract}


\maketitle
\section{INTRODUCTION}
Superconducting thin film $\it RF$/microwave resonators play a
prominent role in many applications including quantum
computing~\cite{Schoelkopf, Bialczak}, single photon
detection~\cite{Goltsman}, bifurcation amplifiers~\cite{Metcalfe}
along with the quest to develop novel devices~\cite{Burke, Mazin}
and media such as metamaterials~\cite{KurterAPL, KurterEIT,
AnlageJOpt}. However, superconductors show nonlinear response when
driven strongly by $\it RF$ signals/microwaves~\cite{Samoilova,
Hein, Ghigo, Ku}, that manifests itself with a significant
dependence of the surface resistance and reactance on the input
power~\cite{Chin, OatesPRL, WosikIEEE, KurterNonlin}, $P_{RF}$. It
is important to find an effective way to keep the resonant
characteristics linear for a long range of $P_{RF}$ to maximize the
power handling capability of the resonators and expand their range
of applicability.

Many superconducting resonators generally employ planar geometries
made up of finite-width thin strips to carry a longitudinal high
frequency current. The magnetic fields generated by flowing currents
along the strips have a common characteristic of being perpendicular
to the edges of the strip. Such a field configuration poses a
challenge to the superconductor. In order to remain in the Meissner
state, the strip must generate strong diamagnetic shielding currents
to screen the perpendicular magnetic field from its interior. This
gives rise to a large current build-up at the edges of a
superconducting film shaped into a strip-geometry
resonator~\cite{RicciIEEE, ZhuravelAPL06}. Screening currents can
approach or exceed the critical current at the edges leading to a
local breakdown of superconductivity and the onset of nonlinear
behavior~\cite{ZhuravelIEEE07}. Therefore, the microwave properties
of superconducting resonators are strongly dependent on the geometry
of the design~\cite{KurterNonlin}.

Apart from simple single strip-lines, co-planar
waveguides~\cite{GhigoAPL, ZhuravelAPL06}, hairpin~\cite{Willemsen}
and meander-line resonators~\cite{ZhuravelJAP} are other planar
designs based on strip geometries. Many of these designs include
parallel conductors where the currents in neighboring strips flow in
opposite directions, see Fig.~1(a). This causes the induced normal
oriented magnetic fields to be enhanced between the strips, and in
turn results in an accumulation of $\it RF$ screening currents at
the edges. Such an inhomogeneous $\it RF$ current density, $J_{RF}$
can create changes in the superconducting properties of the film,
therefore limiting the functionality of the superconducting
resonator by leading to non-linearity in its response even at low
stimulus.

\begin{figure}
\centering
\includegraphics[bb=11 565 569 760,width=3.3 in]{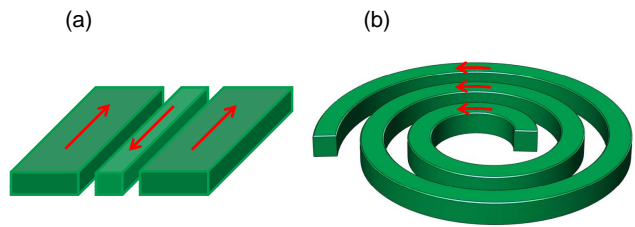}
\caption{(Color online) Schematic sketch of a coplanar wave guide
(a) and a spiral resonator (b). The $\it RF$ currents flow as shown
with the red arrows when the resonators are excited.} \label{Fig1}
\end{figure}

Here we consider a unique resonator in the form of a continuous
planar spiral designed to generate a strong electromagnetic response
below 100 MHz. The spirals are intended to be deep sub-wavelength
meta-atoms of a metamaterial which could be utilized, e.g., in
magnetic resonance imaging applications as compact and low loss flux
guides~\cite{Wiltshire, KurterIEEE}. The resonators have a superior
geometry in which the currents flowing in neighboring strips are in
the same direction and approximately equal in magnitude, at least
for the first few resonant modes, see Fig.~1(b). The perpendicular
components of the induced magnetic fields largely cancel in the
region between the windings, leading to a magnetic field pattern
mainly parallel to the plane of the strips. This renders the
distribution of total current density to be relatively uniform
within the sample compared to the anti-parallel current case
discussed above, eliminating $\it RF$ current build-up at the edges
of the windings. This kind of configuration maintains the linear
characteristics at elevated excitation power, and can be a better
candidate in applications requiring linear $\it RF$/microwave
response.

We have applied the spatially-resolved technique of low-temperature
Laser Scanning Microscopy ($\it LSM$) to map $\it RF$ current
distributions globally (on the entire sample) and locally (in an
individual winding) on spiral resonators made of Nb thin films. From
the two dimensional (2D) $\it LSM$ images of the spirals excited at
the fundamental resonance we have observed an unconventional $\it
RF$ current pattern with the absence of a build-up at the edges of
the turns until a critical power value is reached. The evolution of
the $\it RF$ current distribution with increasing $P_{RF}$ has been
examined to investigate the power handling capability of these
resonators.

The $\it LSM$ technique has various contrast modes for
imaging~\cite{ZhuravelLTP}. Here, we have utilized only two of them:
optical reflectivity and ordinary high-frequency photoresponse
modes. As was demonstrated in previous works~\cite{ZhuravelAPL06,
ZhuravelIEEE07}, the response of the ordinary high-frequency
photoresponse mode is a superposition of two components; inductive
and dissipative/resistive responses. Both generally require the
superconducting sample to show a nonlinear response under laser
irradiation. At low $P_{RF}$ values, only laser heating plays a
significant role in the nonlinearity, however once the power is
elevated, extra dissipation mechanisms will be added due to $\it RF$
heating.  Such a response in superconductors well below their
critical temperature, $T_c$ is mainly attributed to the formation of
local dissipative (non-superconducting) domains where $J_{RF}$ may
exceed the local critical current density, $J_c$. The
superconducting state is extremely sensitive to variations in the
superfluid density that changes either with temperature or magnetic
field, hence nonlinearity is inevitable~\cite{KurterNonlin}. The
effect manifests itself globally as distortion and/or bistable
switching in the resonant transmission as a function of frequency,
$|S_{21}(f)|$, at some microwave powers~\cite{Ghigo, Ku,
KurterNonlin, Brenner} due to increased absorption of microwave
radiation by quasi-particles.

\section{Sample}

The $\it LSM$ measurements presented in this paper use planar spiral
resonators fabricated with 200 nm Nb thin films sputtered onto 350
$\mu m$ thick single crystal quartz substrates. Photolithography and
reactive ion etching ($CF_4$:$O_2$, 90$\%$:10$\%$) are applied to
give a spiral shape to the thin film. The $T_c$ of the Nb film (9.2
K) is obtained from resistance vs temperature
measurements~\cite{KurterIEEE}. Below the $T_\mathrm{c}$ of Nb, the
microwave surface resistance, $R_s$ of the film will be very small
(about \unit{20}{\micro\ohm} at \unit{10}{\giga\hertz} and
\unit{4}{\kelvin})~\cite{Pambianchi94}.

Each spiral is made up of 40 turns, has an outer diameter of 6 mm
and an inner diameter of 4.4 mm. The windings in the spirals and the
spacing between them are of 10 $\mu m$ width.  Prior results show
that the spirals act as very compact self-resonant strips,
supporting up to 10 half-wavelength standing waves of current along
their length~\cite{KurterIEEE}.

\section{RF Excitation}

A single spiral resonator is placed on a sapphire disk plate (50 mm
in diameter, 2 mm in thickness) where a thermometer is attached
nearby, in a cryogenic environment. The sample is stimulated with
$\it RF$ power applied via two coaxial cables terminated by shorted
loops at the end with a diameter slightly larger than the outer
diameter of the spiral as shown in Fig.~2. The planes containing the
excitation ($\it RF$ in) and the pickup loops ($\it RF$ out) are
parallel and the two loops are placed sandwiching the sample between
them~\cite{KurterAPL}. The sample temperature is controlled with a
heater located on the Cu cylinder on the cold head supporting the
sapphire plate. The global resonant response was characterized with
transmission measurements at different $\it RF$ power levels between
$P_{RF}$= -30 dBm and +30 dBm and at a bath temperature of $T_B$=
4.5 K using a Microwave Vector Network Analyzer (Anritsu MS4640A).
From these measurements, the fundamental resonant frequency is found
to be $\sim$74 MHz, followed by higher harmonics.

\section{Cryogenics}

Cooling the spiral samples in the range $T_c$$\geq$$T_B$$\geq$4.5 K
takes place inside the vacuum cavity of a variable temperature
optical cryostat. The temperature of the cold Cu cylinder below the
sample (50 mm in outer diameter with a 5 mm thick wall), see Fig.~2,
is stabilized with an accuracy of 1 mK. The cylinder temperature is
controlled with a bifilar coil heater connected to the temperature
controller and wound around the cold Cu plate having the same
temperature as the cylinder. This Cu cylinder also cools both
coaxial cables to eliminate a possible temperature gradient with the
sample. The top surface of the sample faces the laser probe while
the bottom surface is temperature stabilized by gluing it to the
sapphire disk with cryogenic vacuum grease, assuring a reliable
thermal heat sink. The same grease is used on the thermally
conducting interface between the sapphire and Cu cylinder.

\begin{figure}
\centering
\includegraphics[bb=7 248 598 767,width=3 in]{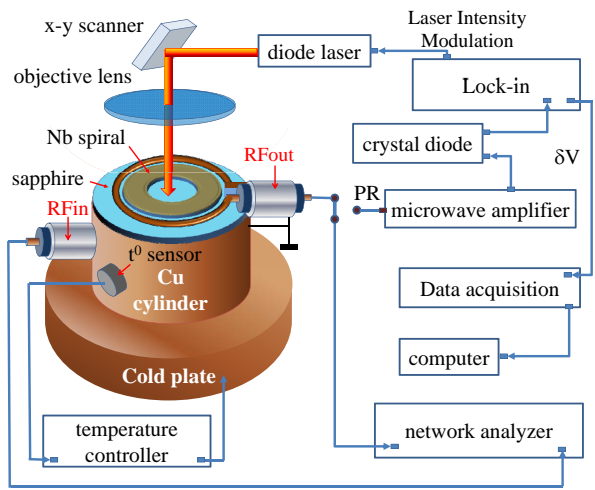}
\caption{(Color online) Simplified schematic representation of the
$\it LSM$ setup used for $2D$ visualization of microwave
photoresponse of the tested resonator structure. Drawing is not to
scale.} \label{Fig2}
\end{figure}

\section{Laser Scanning Microscopy (LSM)}

For $\it LSM$ imaging, the spirals are excited by $\it RF$ signals,
(while being kept well below the $T_c$ of Nb) and illuminated by a
focused laser beam acting as a non-contact optical probe. The $\it
LSM$ photoresponse $\it (PR)$ dominantly comes from
thermally-induced changes in the $\it RF$ transmission
characteristics of the spiral due to absorption of the laser light
with a wavelength of 670 nm.  The smallest diameter of the laser
probe spot is 1.5 $\mu m$ when a 20x magnification (NA=0.42)
objective lens is used for detailed $\it LSM$ imaging (scan area up
to 250x250 $\mu m^2$). Large scale (up to an area of 50x50 $mm^2$)
$\it LSM$ images are acquired with an f-theta objective lens
creating a 20 $\mu m$ diameter laser illuminated spot. The intensity
of the laser is modulated at a frequency of 100 kHz creating an
oscillating thermal and/or optical probe. Changes in $|S_{21}(f)|$
due to the laser heating are synchronously detected with a lock-in
amplifier.

In the bolometric (thermal) regime, the $\it PR$$\sim$($\partial
|S_{21}(f)|$/$\partial T$)$\delta T$ due to local temperature change
$\delta T$, can be uniquely decomposed into inductive and resistive
components.~\cite{ZhuravelAPL06} The inductive photoresponse, $PR_I$
is proportional to $A\lambda^2(x,y) J_{RF}^2(x,y) \delta\lambda$ ,
where $A$ is the area heated by the laser spot and $\lambda(x,y)$ is
the local value of the penetration depth at position $(x,y)$ and can
be interpreted as arising from the changes in penetration depth,
$\delta \lambda$ induced by the laser heating.  When $\lambda(x,y)$
and $\delta \lambda$ have uniform values, the $PR_I$ has a profile
proportional to the local value of $\it RF$ current density squared,
$J_{RF}^2(x,y)$. The resistive photoresponse, $PR_{R}$ arises from
thermally caused changes in the local resistance of the sample $R_s$
and is a convolution of the laser modulated surface resistance,
$\delta R_s$ weighted by the local value of $J_{RF}^2 (x,y)$.

In non-equilibrium (non-thermal) mode, the main mechanism of the
$\it LSM$ $\it PR$ contrast is the following. Below $T_c$, the
absorbed portion of laser power, $\delta P_L$ causes nonequilibrium
changes in the quasiparticle population, $N_{QP}$, resulting from
the high-energy excitation of the superconducting film by individual
optical photons with an energy of $hf_L$= 1.85 eV $\gg$
$2\Delta_{Nb}(0)$ where $\Delta_{Nb}$ is the superconducting energy
gap of Nb, $\it h$ is Planck's constant, and $f_L$ is the laser
(irradiation) frequency. Because of electron-electron and
electron-phonon scattering as well as direct Stokes-like depairing
(with continuous frequency spectrum $hf$$\leq$$hf_L-2\Delta_{Nb}$),
every high-energy quasiparticle is capable of producing an extra
population of low-energy excitations, $N_{QP}$= $\gamma
hf_L$/$2\Delta_{Nb}$, where $\gamma$ is the quantum efficiency and
smaller than 1. The excess quasiparticles create a non-equilibrium
superconducting state due to the reduced superfluid density beneath
the laser probe. As a result, local changes in $N_{QP}(\rho,\delta
P_L)$ cause modifications in the surface impedance $\delta
Z_s(\rho,\delta P_L)$= $\delta R_s(\rho,\delta P_L)+i\omega \delta
L_k(\rho,\delta P_L)$ due to $\delta R_s$ as well as photoinduced
changes in local kinetic inductance, $\delta L_k$ . Here, $\rho$=
$\sqrt{(x-x_0)^2+(y-y_0)^2}$ is the radial coordinate on the sample
surface relative to the position $(x_0,y_0)$ of the laser probe
focus.

We observe that the Nb samples do not show any significant inductive
photoresponse, $PR_I$ at temperatures well below $T_c$. While
increasing both $\it RF$ and/or laser power, it has been found that
resistive photoresponse, $PR_R$ is produced at a lower critical $\it
RF$ power, $P_{c1}$ corresponding to the first local switching of
the sample into the nonlinear regime. The first detectable resistive
component of $\it PR$ can be written as

\begin{equation}
PR_R \propto |S_{21}(f,P_{RF})|^2-|S_{21}(f,P_{RF}+\delta P_L)|^2
\end{equation} for a condition of $J_{RF}$$\geq$$J_c(x_0,y_0,P_{RF})-\delta J_c
(x_0,y_0,P_L)$ combining the effects of the local microwave field (first
term) and suppression of the critical current by the laser beam
(second term).

It has been shown in the literature (see, for instance,
Refs.~\cite{Zharov, KurterNonlin}) that the first nonlinear
distortion of $|S_{21}(f)|$ appears as a deviation where the
$|S_{21}(f)|$ curves fall on to curves with smaller quality factor,
$\it Q$, in a narrow-band near the resonant frequency $f_0$ (with
$P_{RF}$ exceeding $P_{c1}$). In the case of a small optical probe
perturbation $\delta P_L$$\ll$$P_{c1}-\delta P_L$, the resistive
component of $\it LSM$ $\it PR$ may be undetectable outside this
narrow band, while strong $\it PR$ signals are generated inside the
band.

The $\it LSM$ work presented here follows a modified procedure
originally developed in Ref.~\cite{ZhuravelIEEE07} which is based on
the insertion loss component of the photoresponse, $PR_{IL}$, rather
than $PR_I$ and $PR_R$ measured at a frequency in the vicinity of
$f_0$. At a fixed $\it RF$ frequency and spatially independent laser
probe perturbation, the $\it LSM$ $\it PR$ is proportional to the
laser-beam-induced changes in resonator transmission, $\delta
|S_{21}(f)|^2$ that can be expressed in a form close to that
introduced in Ref.~\cite{ZhuravelMSMW}.

\begin{eqnarray}\nonumber
PR \propto \delta |S_{21}(f)|^2 = \frac{1}{2} (\frac{\partial
|S_{21}(f)|^2}{\partial f_0} \frac{\partial f_0}{\partial
P_L}+\\\frac{\partial |S_{21}(f)|^2}{\partial
(1/2Q)}\frac{\partial (1/2Q)}{\partial
P_L}+\frac{\partial|S_{21}(f)|^2}{\partial \hat{S_{21}}^2}
\frac{\partial \hat{S_{21}}^2}{\partial P_L}) \delta P_L
\end{eqnarray} where the transmission coefficient, $|S_{21}(f)|^2$ [ratio
of the transmitted power, $P_{RF}^{OUT}(f)$, to the input power,
$P_{RF}(f)$] as a function of driving frequency $f \sim f_0$ is
given in the limit of weak coupling by~\cite{Petersan}

\begin{equation}
|S_{21}(f)|^2=\frac{\hat{S_{21}}^2}{1+4Q^2(f/f_0-1)^2}
\end{equation}
and $\hat{S_{21}}^2$ is the maximum of the transmission coefficient
at the peak of the resonance. By substitution of Eq.~3 in Eq.~2, one
finds that the inductive

\begin{equation}
PR_I \propto \frac{\partial |S_{21}(f)|^2}{\partial
f_0}=\frac{8\hat{S_{21}}^2
Q^2(\frac{f}{f_0}-1)}{[1+4Q^2(\frac{f}{f_0}-1)^2]^2}
\frac{f}{f_{0}^2}
\end{equation} and the resistive

\begin{equation}
PR_R \propto \frac{\partial
|S_{21}(f)|^2}{\partial(1/2Q)}=\frac{16\hat{S_{21}}^2
Q^3(\frac{f}{f_0}-1)^2}{[1+4Q^2(\frac{f}{f_0}-1)^2]^2}
\end{equation} components of total $\it LSM$ $\it PR$ are nulled at $f=f_0$, while the insertion loss component
\begin{equation}
PR_{IL} \propto
\frac{|S_{21}(f)|^2}{\delta(\hat{S_{21}})}=\frac{2\hat{S_{21}}}{1+4Q^2(\frac{f}{f_0}-1)^2}
\end{equation} is peaked at  $f=f_0$.

In terms of local photo induced changes, $PR_{IL}$ is directly
linked with Ohmic dissipation generated by the laser probe at
position $(x_0, y_0)$,~\cite{ZhuravelMSMW2010, Culb}
\begin{equation}
PR_{IL}(x_0,y_0) \propto J^2_{RF}(x_0,y_0)\delta R_s(x_0,y_0)
\end{equation}

In the frame of the paradigm described in Ref.~\cite{ZhuravelIEEE07}
(in the case of a linear response function and a small probe
perturbation) for a strip geometry oriented along the path $L$ in
the $\ell$ direction, the change in surface resistance due to a
change in local critical current
[$J_{RF}$$\geq$$J_c(\ell_0,P_{RF})-\delta J_c (\ell_0,P_L)$] at a
specific laser probe position $\ell_0$ may be described as

\begin{eqnarray}\nonumber
\delta R_s(\ell_0) \propto \frac{\pi}{4} \frac{\Lambda}{WL}
\int_{\L} d\ell \frac{\partial R_s(\ell_0)}{\partial
J_c(\ell_0)}|_{J=J_{RF}}
\\\frac{\partial J_c(\ell_0)}{\partial P_L}|_{P=P_{RF}^{CIRC}+\delta
P_L} \delta P_L(\ell_0)
\end{eqnarray} for large scale imaging mode ($\Lambda$$\geq$$W$) where $\L$ is the path along the entire spiral with total length of $L$, $W$ is the width of the film,
$P_{RF}^{CIRC}$ is the circulating $\it RF$ power in the resonator
and $\Lambda$ is the characteristic healing length describing
spatial decay of $PR_{IL}(\ell_0)$ $\propto$ $e^{-|\ell -
\ell_0|/\Lambda}$ at a distance $\ell$ outside the intense beam
focus.

As was postulated in Ref~\cite{Gross}, one can assume that both
quantities $\partial R_s(x_0)$/$\partial J_c(x_0)$ and $\partial
J_c(x_0)$/$\partial P_L$ are invariable in the probed sample area if
$\it d$, $\delta P_L$ and $\Lambda$ are spatially uniform through
the whole resonator structure. Combining Eq.~7 with the integral
value of Eq.~8 over the laser probe profile $\Lambda \delta P_L$
leads to
\begin{eqnarray}\nonumber
PR_{IL}(x_0) \propto \frac{1}{Wd} J_{RF}^2(x_0)\frac{\partial
R_s(x_0)}{\partial J_c(x_0)}|_{J=J_{RF}}\\\frac{\partial
J_c(x_0)}{\partial P_L}|_{P=P_{RF}^{CIRC}+\delta P_L} d \Lambda
\delta P_L(x_0)
\end{eqnarray} at location $x_0$ in the one dimensional strip geometry.

Note that Eq.~9 demonstrates a threshold mechanism of $PR_{IL}$
generation relative to excitations by both $P_{RF}^{CIRC}$ and
$\delta P_L$. In the undercritical state of the superconducting
structure at $P=P_{RF}^{CIRC}+\delta P_L \leq P_{c1}$, the value of
$\partial R_s(x_0)$/$\partial J_c(x_0)$ is zero at any position of
the laser probe. In this case there is no $PR_{IL}(x_0)$ detectable
by the $\it LSM$ technique at $f_0$ in microwave imaging mode.  In
addition, very weak response is observed in purely normal regions of
the sample. A detectable $PR_{IL}$ signal is generated only in the
narrow range of power between $P_{c1}$ and $P_{c2}$ (upper critical
$\it RF$ power). Note that $P_{c1}$ (see Fig.~4b) denotes the total
($P_{RF}^{CIRC}+\delta P_L $) power initiating the first local
dissipative source that destroys superconductivity. By $P_{c2}$ we
denote the power of this source giving rise to normal state
switching. As seen from Eq.~9, $PR_{IL}(x_0)$ is proportional to
$J^2_{RF}(x_0)$ in this range and spatial variations of $\it LSM$
$\it PR$ amplitude directly show the distribution of $J^2_{RF}(x_0)$
along that part of the standing wave that generates an overcritical
state in the superconducting strips. Any deviation of $PR_{IL}(x_0)$
from the shape of a sinusoidal standing wave pattern then gives
evidence for an inhomogeneous distribution of $J_c(x_0)$ due to the
term $\partial J_c(x_0)$/$\partial P_L$ in Eq.~9. Also, it is clear
that manipulations by both $P_{RF}$ and $\delta P_L$ may be used to
probe local values of $J_c(x_0)$ as either $P_{RF}$ or $\delta P_L$
is increased.

In the case of 2D $\it LSM$ probing (characteristic length of the
laser-probe induced non-equilibrium state, $\Lambda$ $\leq$ strip
width, W), the main $\it LSM$ $\it PR_{IL}$ imaging mode results
from laser probe induced redistribution of the microwave current
around the illuminated area. This effect leads to additional Ohmic
dissipation in the nearby unilluminated areas of the superconducting
strip generating

\begin{eqnarray}\nonumber
PR_{IL}(x_0, y_0) \propto \frac{\Lambda^2}{WL}<J_{RF}^2>_{W-W_c}
\\\frac{\partial R_s(x_0,y_0)}{\partial J_c(x_0,y_0)}\frac{\partial J_c(x_0,y_0)}{\partial P_L}
 \delta P_L
\end{eqnarray}
The effect is linked with the laser-induced modulation of the local
critical current,
\begin{eqnarray}
\delta I_{c}(x_0, y_0)=\frac{\pi \Lambda^2}{4}\frac{\partial J_c(x_0,y_0)}{\partial P_L}
 \delta P_L
\end{eqnarray}
underneath the laser probe allowing direct measurement of
$I_c$. Here, $\partial J_c(x_0,y_0)$/$\partial
P_L$$\propto$$J_c(x_0,y_0)$ if $\Lambda$ and $\delta P_L$ are
independent of the beam position. Larger critical current densities
produce larger $\it LSM$ $PR_{IL}$ as a result of redistribution of
$J_{RF}$ through the cross-section of the undercritical currents
of width $W-W_c$, thus increasing the averaged $J_{RF}$ flowing
there. Here $W_c$ denotes the width of the critical region.

As one can see from Eq.~10, the highest microwave current densities
produce the largest  $PR_{IL}(x_0, y_0)$ resulting in quantitative
profiles of $J_{RF}^2 (x_0,y_0)$ in the area of the laser beam
raster on the superconductor surface.

\section{GLOBAL PHOTORESPONSE RESULTS}

To characterize the resilience of the superconducting spiral
resonators at high $P_{RF}$, it is important to examine how current
is distributed in the entire sample when driven by strong $\it RF$
signals. Fig.~3(a) is a 2D $\it LSM$ image showing the global
photoresponse of a Nb spiral excited at its fundamental resonant
mode of 74 MHz. The laser is scanned over a 7.6 x 7.6 $mm^2$ area at
$T_B$= 4.5 K, $P_{RF}$= 14.8 dBm and 1 $mW$ laser power. The
contrast in the image is mainly produced by $PR_{IL}$ where the
bright areas can be interpreted as $J_{RF}^2(x,y)$ to first
approximation, and illustrates a mode in which a single
half-wavelength of standing wave current spans the length of the
spiral.  As seen, $\it RF$ current mainly flows in the middle
windings in a quite uniform way, and diminishes towards the inner
and outer edges of the spiral. Fig.~3(b) is a 2D $\it LSM$
reflectivity image of the same spiral and shows the turns in an area
on the spiral shown with the green box in (a).

The evolution of the $PR_{IL}$ coming from the individual windings
along the cross section of the spiral [marked as S in (a)] is shown
in Fig.~3(c) for four different $P_{RF}$ values and reveals the
$J_{RF}$ distribution in greater detail; note that the maximum $\it
PR$ corresponds to the center of the S cut line, and the ends show
no response, confirming what is seen in (a). The asymmetric shape of
the standing wave profile in the fundamental mode is understood from
the fact that the spiral turns at larger radius have a greater
circumference.  The dots show the estimated $J_{RF}^2$ along the cut
$S$ for the case of a half sinusoid wave wrapped into a spiral.
These dots describe well the observed $PR_{IL}$, indicating that the
measured $\it PR$ distribution is quite similar to the naive
interpretation of imaging $J_{RF}^2(x,y)$.

In Fig.~4(a), the individual line-scans of $PR_{IL}$ at different
incident power levels are shown in a three dimensional (3D) image.
Fig.~4(b) shows power-dependent evolution of $\it LSM$ $\it PR$ at
three fixed positions of the laser probe coinciding with the centers
of three neighboring Nb strips (strips A, B and C) exposed to
maximum $J_{RF}$ near the peak of the microwave standing wave. Note
that a linear power scale is used. In the purely superconducting or
normal states, $\it LSM$ $\it PR$ is not observable [notice the zero
$\it PR$ at the low and high limits of $P_{RF}$ values in (a)]. As
evident from Fig.~4(b), $PR_{IL}(S_{A,B,C}, P_{RF}) \propto
J_{RF}^2(\sim P_{RF})$ shows an almost linear trend for a long range
of nonequilibrium states of the Nb film starting from an $P_{RF}$
corresponding to the first observable $\it LSM$ $\it PR$ at $P_{c1}$
up to a switching to the normal state at $P_{c2}$= 14.8 dBm where
the $\it PR$ drops. These observations validate Eq.~9 in explaining
our results.  Also, one can see that based on the values of
$P_{c1}$, the $J_c$ of all three strips is practically the same,
indicating a spatial uniformity in Nb film microstructure.

\begin{figure}
\centering
\includegraphics[bb=11 187 598 619,width=3 in]{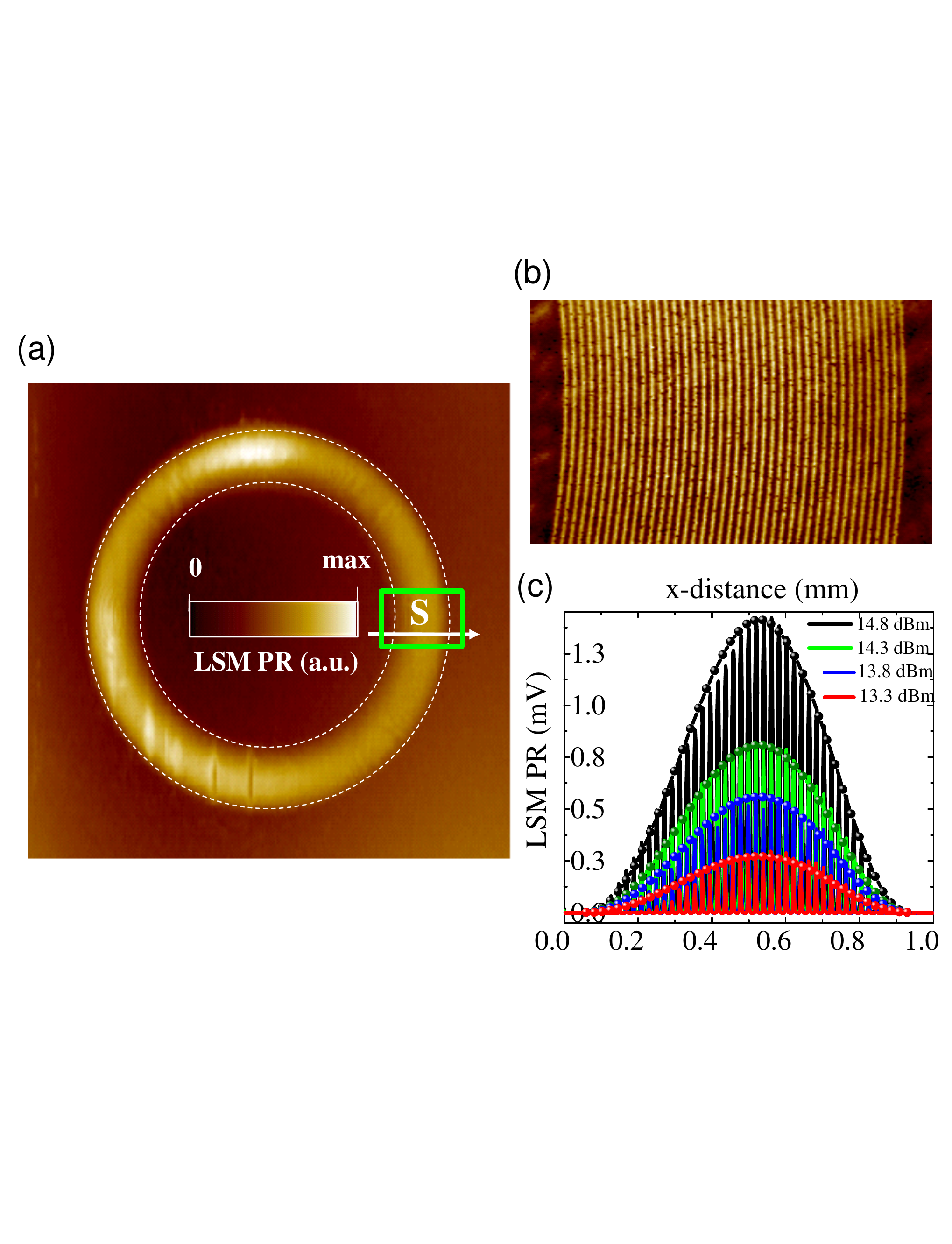}
\caption{(Color online) (a) 2D $\it LSM$ image showing current
distributions in a Nb spiral with an outer diameter of 6 mm and 40
turns, at the fundamental resonant mode of 74 MHz, $T_B$= 4.5 K,
$P_{RF}$= 14.8 dBm. (b) 2D $\it LSM$ reflectivity image showing the
individual turns within an area on the spiral marked with a green
box in (a). (c) The power dependent $PR_R$ along the cross section
of the spiral shown with $\it S$-line; maximum at the center,
minimum at the edges. The dots are the estimated $J^2_{RF}$ profile
for a simple standing wave current pattern at each $P_{RF}$.}
\label{Fig3}
\end{figure}

\begin{figure}
\centering
\includegraphics[bb=22 508 583 756,width=3.3 in]{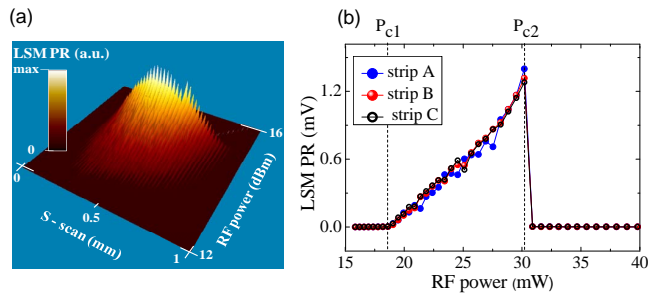}
\caption{(Color online) (a) 3D $\it LSM$ image showing the power
dependence of $PR_{IL}$ over the S-line scan shown in Fig.~3(a). (b)
Experimental $\it LSM$ $\it PR$  vs. $P_{RF}$ on a linear scale,
taken at three neighboring strips (strip~B is located at the center
of the S-line scan between strip~A and strip~C). Both data are
obtained at a temperature well below $T_c$, 4.5 K.} \label{Fig4}
\end{figure}

\section{LOCAL PHOTORESPONSE RESULTS}

Upon more detailed examination of the Nb resonators, one can see
that the $\it LSM$ $\it PR$ is also the strongest at the center of
an individual turn forming the spiral, following the same trend of
the global behavior shown in Fig.~3. Fig.~5(a) shows 2D $\it LSM$
$PR_{IL}$ of a 40x40 $\mu m^2$ area on the same resonator scanned
with a 1.5 $\mu m$ diameter laser probe in the fundamental resonant
mode of the spiral, a $T_B$ of 4.5 K, and $P_{RF}$ of 7 dBm, and
laser power of 1 $\mu W$, while Fig.~5(b) is a $\it LSM$
reflectivity image obtained from the same area. By comparing the
simultaneously measured $\it LSM$ $\it PR$ and reflectivity one
finds that the $\it PR$ is centered in the windings and does not
extend to the edge of the strip. This observation is verified by
studying the $\it RF$ $\it PR$ images as a function of increasing
temperature or $P_{RF}$, and noting that the $\it PR$ spreads out
laterally in both directions from the center of the strip as the
critical temperature and power are approached (see the power
evolution of $\it PR$ coming from Nb turns in Fig.~5(c); beyond +10
dBm the entire strip starts to show strong resistive response).

The accumulation of $\it PR$ in the center of the turns is in
contrast with previously published $\it PR$ profiles of
strip-geometries that show substantial concentration of the $\it
LSM$ $\it PR$ at the edges of current-carrying
strips~\cite{ZhuravelAPL02, Culb, ZhuravelJSNM, ZhuravelLTP}.
Qualitatively, this fact can be easily understood. Here, the
vertical components of magnetic field between the strips are
partially cancelled, as discussed above, since adjacent strips have
nearly equal and parallel currents (at least in the fundamental
mode). The spiral effectively acts like a disk carrying an
approximately homogeneous current distribution, in which the current
density goes to zero at the inner and outer radii of the disk.

As well as $\it P_{RF}$, laser power has an impact on the $\it RF$
$\it PR$ profile in the spirals. Fig.~5(d) shows the initial
depression of $J_c(x_0,y_0)$ by modulated laser power, $\delta
P_L(x_0,y_0)$= 1 $\mu W$ in detail (blue curve). The main feature of
$PR_{IL}(x_0,y_0)$ induced by $\delta P_L$ is generated only inside
a very narrow resistive strip, directed along the center of the
strip-line. Moreover, no spatial modulation in $\it LSM$ $\it PR$ is
visible in the scanned area along the direction of $\it RF$ current
flow, indicating that the Nb film is quite homogenous, which rules
out a structure-related mechanism of hot-spot formation. Taking
Eqs.~1 and 10 into account, as well as considering the fact that the
laser beam illumination is spatially uniform, one can deduce that
$\it RF$ current is peaked half way between the Nb strip-line edges
reaching local maxima of $J_{RF}(x_0,y_0)$$\leq$$J_c(x_0,y_0)$
there. Larger laser power (see the red curve in Fig.~5(d)
corresponding to $\delta P_L(x_0,y_0)$= 10 $\mu W$) increases the
area of the strip in the critical state and consequently
$J_{RF}(x_0,y_0)$ adjusts itself accordingly, since in the
superconducting state, $J_{RF}(x_0,y_0)$ cannot exceed
$J_c(x_0,y_0)$. Thus, the distribution of $PR_{IL}(x_0,y_0)$ spreads
all over the strip occupying the dissipative regions of the still
superconducting strip.

Line-scan profiles across two strips of such spatial evolution of
$J_c(x)$ are shown in Fig.~6 as a function of $P_{RF}$. Small laser
probe perturbation ($P_L$= 1 $\mu$W $<<$$P_{c1}$) generates the
first observable $\it LSM$ $PR_{IL}$ exactly at the centers of the
superconducting strips carrying a current density $J_{RF}(x_0,
y_0)$= $J_c(x_0, y_0)$ at $P_{RF}$ = $P_{c1}$ = 12.8 dBm, described
by Eq.~10. Emergence of this signal is linked with the creation of
sub-micron critical-state nonequilibrium domains at the centers of
the strips, much smaller than the size of the laser probe. The full
width half maximum (FWHM) of the position dependent $\it LSM$ $\it
PR$ is about 2$\Lambda$ (see Fig.~5d and Fig.~6). Further increase
in $P_{RF}$ leads to a broadening of the critical state area which
results in an increase of the FWHM of the dissipative $\it LSM$
$PR_{IL}$ profiles. With reference to Eq.~10 we see that as $J_{RF}$
increases, the width of the film in the critical state ($W_c$) will
increase, forcing more current into the under-critical region
($W-W_c$) and thus increasing $<J_{RF}^2>_{W-W_c}$.

\begin{figure}
\centering
\includegraphics[bb=14 155 598 724,width=3 in]{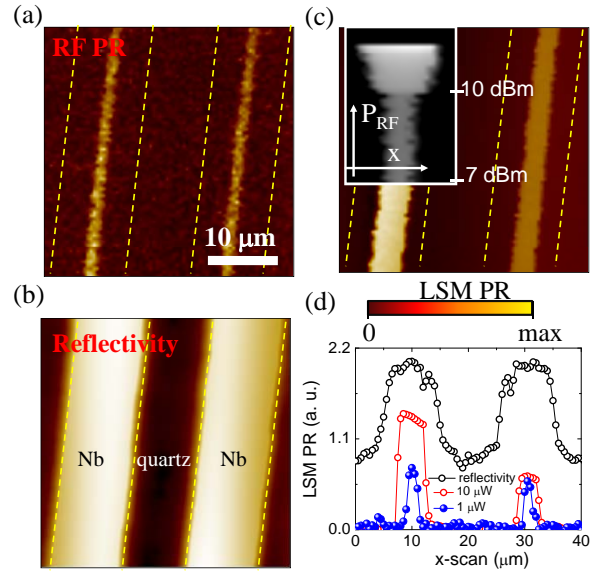}
\caption{(Color online) 2D $\it LSM$ (a) $PR_{IL}$ and (b)
reflectivity images taken from 40x40 $\mu m^2$ area on  the Nb
spiral resonator at a laser power of about 1 $\mu W$. (c) 2D $\it
LSM$ $PR_{IL}$ image at 10 $\mu W$. Inset shows $\it RF$ power
dependence of $\it LSM$ $PR_{IL}$ on the same area, showing the
$J_{RF}$ profiles at low and high $\it RF$ stimulus.  The x-line cut
is at the same location in the figure and inset. (d) $\it LSM$ $\it
PR$ coming from 2 neighboring Nb turns at two different laser
powers; 1 $\mu W$ and 10 $\mu W$. The data are taken at $P_{RF}$= 7
dBm.} \label{Fig5}
\end{figure}

Further, we estimate $J_c$ by using the measured $\it LSM$
$PR_{IL}$. In the inset of Fig.~5(c), there is a sharp transition
from a center concentrated strip-like resistive state to an almost
uniform resistive state where $\it PR$ covers the whole width of the
strip. For a wide range of $P_{RF}$ below $P_{c1}$$\sim$ 10 dBm, the
$\it LSM$ $\it PR$ is almost independent of $\it RF$ stimulus. We
associate this effect with the auto-adjusting of instantaneous
circulating $J_{RF}$ to a value which could be accommodated in the
superconducting resonator with varying local values of $J_c$.
Considering the absence of visible imperfections in the reflectivity
data taken from the same area of this spatial power dependence we
can confirm that this transition results from heating effects
generated by hot-spot formation at a location far from the scanned
line. Thus, this defect-free section of the scan is chosen for rough
estimation of the $J_c(x_0,y_0)$ limit for our resonator. Estimation
is done based on measurements of $P_{c1}$ generating the first
detectable $\it LSM$ resistive image similar to that shown in
Fig.~5(a).

\begin{figure}
\centering
\includegraphics[bb=11 227 594 666,width=3 in]{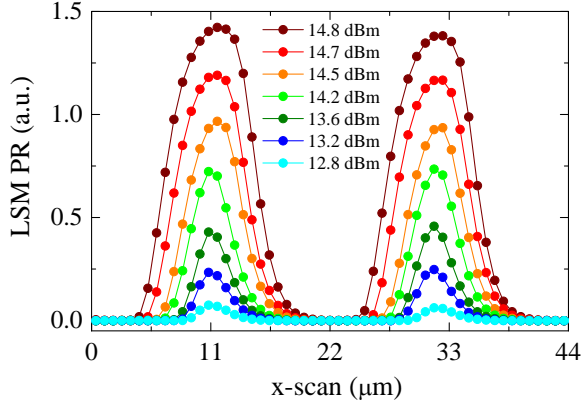}
\caption{(Color online) Power dependence of dissipative $\it LSM$
$PR_{IL}$ profiles showing the broadening of the critical state
along the same 40 $\mu m$ line x-scan through the width of two Nb
turns at a laser power of about 1 $\mu W$.} \label{Fig6}
\end{figure}

Since the use of smaller laser power is favorable in these
measurements in terms of eliminating extra dissipation, we have
scanned the spiral surface at $\delta P_L$ = 500 nW and found that
$P_{c1}$$\approx$9.5 dBm. There is no $\it LSM$ $\it PR$ observed
below this laser power except only at very high $P_{RF}$. Using
measured data [$P_{c1}$= 8.91 mW (9.5 dBm), $W$= 10 $\mu m$, $d$=
0.2 $\mu m$, $Q$= 545.4, $|S_{21}(f_0)|$= 0.1553, harmonic number
$n$ = 3 and $Z_0$= 96 $\ohm$ which is the characteristic impedance
estimated for a co-planar waveguide of similar geometry] and a
simple model for homogenous current distribution in the stripline we
estimate an upper limit of $J_c$ as~\cite{Oates}
\begin{equation}
J_c(f_0)=\frac{1}{Wd} \sqrt{\frac{S_{21}(f_0)(1-S_{21}(f_0))4QP_{c1}}{n
\pi Z_0}}
\end{equation}
and obtain  $J_c$$\sim$ 2.7$\times$$10^{10}$ $A/m^2$  from Eq.~12.
This value is more than an order of magnitude smaller than the
theoretical estimation for the depairing current density
$J^{GL}_{dp}(T/T_c)$$\sim$$J^{GL}_{dp}(0)(1-T/T_c)^{3/2}$, which is
0.44$\times$$10^{12}$ $A/m^2$ at 4.5 K (using $J^{GL}_{dp}(0)$ =
1.26$\times$$10^{12}$ $A/m^2$ for Nb at $T_B$= 0 K~\cite{Yu}).  This
result implies that the measured critical current is limited by
factors other than the depairing limit.

\section{HIGHER HARMONICS}

We observe that higher order harmonic modes of the Nb spiral
resonator have more inhomogeneous current distributions in the
windings due to the larger spatial gradients of the current.
Moreover, the resonant characteristics of the Nb spiral are more
power dependent in those higher modes than the fundamental mode.

\begin{figure}
\centering
\includegraphics[bb=70 370 486 745,width=3 in]{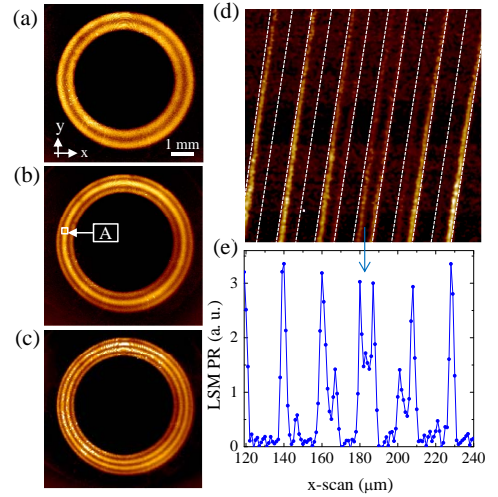}
\caption{(Color online) Large-scale 7x7 $mm^2$ $\it LSM$ $\it PR$
images showing $\it RF$ current induced dissipation in a Nb spiral
at 4.5 K, 10 dBm and at (a) the second harmonic (219 MHz), (b) the
third harmonic (355 MHz), and (c) the forth harmonic (498 MHz). The
area marked A in (b) indicates the position of a detailed 2D image
(d) that shows the same $\it RF$ dissipation in a 125x125 $\mu m^2$
region localized at the center of the $3 \lambda/2$ standing wave
pattern.  (e) $\it LSM$ $\it PR$ profile along a 125 $\mu m$ x-scan
corresponding to the bottom line-scan in (d).} \label{Fig7}
\end{figure}

Fig.~7(a)-(c) show $\it LSM$ $\it PR$ in the $2^{nd}$ to $4^{th}$
harmonic standing wave patterns over a 7x7 $mm^2$ area. If we focus
on a small area (125x125 $\mu m^2$) in the 3rd harmonic $\it LSM$
image shown in (b), we see a different scenario from that discussed
above, Fig.~7(d). In locations where there is a gradient in the
current in the radial direction, a more traditional current profile
is observed (notice the $4^{th}$ turn from left in d). Fig.~7(e)
shows the profile of this $\it PR$ as a function of position in the
radial direction. The PR is uniformly distributed across the strip
at the peak of the standing wave pattern.  However in regions where
there is a significant change in the amount of current flowing in
neighboring strips, the $\it PR$ tends to be concentrated along one
edge of the strip. For example on the right hand side of Fig.~7(d)
there are decreasing magnitudes of current flowing through the
strips to the right. This leads to an asymmetry of the perpendicular
magnetic field so that there is a larger field on the left side of
each gap compared to the right. This in turn leads to an asymmetric
buildup of current on the right side of each strip to screen out the
field. The analogous phenomenon occurs on the other side of the peak
in the current distribution as the current carried by windings
further to the left decrease in magnitude.

We have seen similar effects in the third harmonic standing wave
pattern of similar spirals implemented with $YBa_2Cu_3O_7$.  Two
distinct photoresponse peaks are seen on the edges of the
$YBa_2Cu_3O_7$ strip.  The maximum of the standing wave pattern is
visible in the middle strip, and the current decreases to either
side.

\section{CONCLUSIONS}

In conclusion, we have mapped the global and local current profile
in planar spiral resonators implemented with superconducting Nb thin
films via $\it LSM$ imaging. The PR analyses reveal that the $\it
RF$ current in the fundamental mode mainly flows at the center of
the turns of the spiral, which is contrary to the profile that is
traditionally seen in stripline and coplanar waveguide resonators.
The continuous spiral geometry plays an important role in this
unusual current profile contrasting with the conventional strip
resonator case having anti-parallel currents in adjacent elements
where $\it RF$ current accumulates at the edges.

\vskip 0.5cm
\begin{center}
ACKNOWLEDGEMENTS
\end{center}
We gratefully acknowledge the contributions of John Abrahams, Tian
Lan, Liza Sarytchev, Brian Straughn, and Frederic Sirois. The work
at Maryland was supported by ONR Grants No. N000140811058 and No.
20101144225000, the US DOE DESC 0004950, the ONR/University of
Maryland AppEl Center, Task D10 (N000140911190), and Center for
Nanophysics and Advanced Materials (CNAM). The work in Karlsruhe is
supported by the Fundamental Researches State Fund of Ukraine and
the German Federal Ministry of Education and Research under Grant
No. UKR08/011, the Deutsche Forschungsgemeinschaft (DFG) and the
State of Baden-W¨urttemberg through the DFG Center for Functional
Nanostructures (CFN), and a National Academy of Sciences of Ukraine
program on Nanostructures, Materials and Technologies. S.M.A.
acknowledges sabbatical support from the CFN at Karlsruhe Institute
of Technology.


\begin{thebibliography}{22}

\bibitem{Schoelkopf} R. J. Schoelkopf and S.M. Girvin, Nature {\bf 451}, 664 (2008).
\bibitem{Bialczak} R. C. Bialczak, M. Ansmann, M. Hofheinz, M. Lenander, E. Lucero,
M. Neeley,  A. D. O'Connell, D. Sank,  H. Wang, M. Weides, J.
Wenner, T. Yamamoto, A. N. Cleland, and J. M. Martinis, Phys. Rev.
Lett. {\bf 106}, 060501 (2011).
\bibitem{Goltsman} G. Goltsman, A. Korneev, A. Divochiy, O. Minaeva, M. Tarkhov, N.
Kaurova, V. Seleznev, B. Voronov, O. Okunev, A. Antipov, K. Smirnov,
Yu. Vachtomin, I. Milostnaya, and G. Chulkova, J. Mod. Opt. {\bf
56}, 1670 (2009).
\bibitem{Metcalfe} M. Metcalfe, E. Boaknin, V. Manucharyan, R. Vijay, I. Siddiqi, C. Rigetti, L.
Frunzio, R. J. Schoelkopf, and M. H. Devoret, Phys. Rev. B {\bf 76},
174516 (2007).
\bibitem{Burke} P. J. Burke, R. J. Schoelkopf, D. E. Prober, A. Skalare, W. R. McGrath, B. Bumble, and H. G. LeDuc ,
Appl. Phys. Lett. {\bf 68}, 3344 (1996).
\bibitem{Mazin} Benjamin A. Mazin, Daniel Sank, Sean McHugh, Erik A. Lucero, Andrew Merrill, Jiansong Gao, David Pappas, David Moore,
and Jonas Zmuidzinas, Appl. Phys. Lett. {\bf 96}, 102504 (2010).
\bibitem{KurterAPL} Cihan Kurter, John Abrahams, and Steven M. Anlage, Appl. Phys. Lett. {\bf 96}, 253504 (2010).
\bibitem{KurterEIT} Cihan Kurter, Philippe Tassin, Lei Zhang, Thomas Koschny, Alexander P.
Zhuravel, Alexey V. Ustinov, Steven M. Anlage, and Costas M.
Soukoulis , Phys. Rev. Lett. {\bf 107}, 043901 (2011).
\bibitem{AnlageJOpt} S. M. Anlage, J. Opt. {\bf 13}, 024001 (2011).
\bibitem{Samoilova} T. B. Samoilova, Supercond. Sci. Tech. {\bf 8}, 259 (1995).
\bibitem{Hein} M. Hein, {\it High-temperature-superconductor thin films at microwave
frequencies} (Springer-Verlag, Berlin, 1999).
\bibitem{Ghigo} G. Ghigo, R. Gerbaldo, L. Gozzelino, F. Laviano, E. Mezzetti , Phys. Rev. B {\bf 82}, 054520 (2010).
\bibitem{Ku} J. Ku, V. Manucharyan, and A. Bezryadin , Phys. Rev. B {\bf 82}, 134518 (2010).
\bibitem{Chin} C. C. Chin, D. E. Oates, G. Dresselhaus and M. S. Dresselhaus, Phys Rev B {\bf 45}, 4788 (1992).
\bibitem{OatesPRL} D. E. Oates, S. H. Park and G. Koren, Phys. Rev. Lett. {\bf 93}, 197001 (2004).
\bibitem{WosikIEEE} J. Wosik, L.- M. Xie, R. Grabovickic, T. Hogan, and S. A. Long , IEEE Trans. Appl. Supercond. {\bf 9}, 2456 (1999).
\bibitem{KurterNonlin} C. Kurter, A. P. Zhuravel, A. V. Ustinov, S. M.
Anlage, Phys. Rev. B {\bf 84}, 104515 (2011).
\bibitem{RicciIEEE} M. Ricci, H. Xu, R. Prozorov, A. P. Zhuravel, A. V. Ustinov, S. M. Anlage , IEEE Trans. Appl. Supercond. {\bf 17}, 918 (2007).
\bibitem{ZhuravelAPL06} A. P. Zhuravel, S. M. Anlage, A. V. Ustinov, Appl. Phys. Lett. {\bf 88}, 212503 (2006).
\bibitem{ZhuravelIEEE07} A. P. Zhuravel, S. M. Anlage, A. V. Ustinov, IEEE Trans. Appl. Supercond. {\bf 17}, 902 (2007).
\bibitem{GhigoAPL} G. Ghigo, R. Gerbaldo, L. Gozzelino, F. Laviano, G. Lopardo, E. Monticone, C. Portesi, and E. Mezzetti, Appl. Phys. Lett. {\bf 94}, 052505 (2009).
\bibitem{Willemsen} B. A. Willemsen, T. Dahm, and D. J. Scalapino, Appl. Phys. Lett. {\bf 71}, 3898
(1997); T. Dahm and D. J. Scalapino, J. Appl. Phys. {\bf 82}, 464
(1997).
\bibitem{ZhuravelJAP} A. P. Zhuravel, S. M. Anlage, S. K. Remillard, A. V. Lukashenko, and A. V. Ustinov, J. Appl. Phys. {\bf 108}, 033920 (2010).
\bibitem{Wiltshire} M. C. K. Wiltshire, J. B. Pendry, I. R. Young, D. J. Larkman, D. J. Gilderdale, and J. V. Hajnal,
Science {\bf 291}, 849 (2001).
\bibitem{KurterIEEE} C. Kurter, A. P. Zhuravel, J. Abrahams, C. L. Bennett, A. V. Ustinov, and S. M.  Anlage, IEEE Trans. Appl. Supercond.
{\bf 21}, 709 (2011).
\bibitem{ZhuravelLTP} A. P. Zhuravel, A. G. Sivakov, O. G. Turutanov, A. N. Omelyanchouk, S. M. Anlage, A. Lukashenko, A. V. Ustinov, and D. Abraimov,
Low Temp. Phys. {\bf 32}, 592 (2006).
\bibitem{Brenner}  Matthew W. Brenner, Sarang Gopalakrishnan, Jaseung Ku, Timothy J. McArdle, James N. Eckstein, Nayana Shah, Paul M. Goldbart,
Alexey Bezryadin, Phys. Rev. B {\bf 83}, 184503 (2011).
\bibitem{Pambianchi94} M. S. Pambianchi, S. M. Anlage, E. S. Hellman,
J. E. H. Hartford, M. Bruns, and S. Y. Lee, Appl. Phys. Lett. {\bf
64}, 244 (1994); M. S. Pambianchi, L. Chen, and S. M. Anlage, Phys.
Rev. B {\bf 54}, 3508 (1996).
\bibitem{Zharov} A. A. Zharov, and A. N. Reznik, Technical Physics {\bf 43}, 117 (1998).
\bibitem{ZhuravelMSMW} A. P. Zhuravel, S. M. Anlage, S. Remillard, A. V. Ustinov, Proceedings of the Sixth International Symposium on Physics and
Engineering of Microwaves, Millimeter and Sub-millimeter Waves,
(IEEE, 2007), Vol. 1, pp. 404–406.
\bibitem{Petersan} P. J. Petersan and S. M. Anlage, J. Appl. Phys. {\bf 84}, 3392 (1998).
\bibitem{ZhuravelMSMW2010} A. P. Zhuravel, S. M. Anlage, and A. V. Ustinov, Proceedings of the
Seventh International Symposium on Physics and Engineering of
Microwaves, Millimeter and Sub-millimeter Waves (IEEE, 2010), Vol.
1, pp. 1–3.
\bibitem{Culb} J. C. Culbertson, H. S. Newman, and C. Wilker, J. Appl. Phys. {\bf 84}, 2768 (1998).
\bibitem{Gross} R. Gross and D. Koelle, Rep. Prog. Phys. {\bf 57}, 651 (1994).
\bibitem{ZhuravelAPL02} A. P. Zhuravel, A. V. Ustinov, K. S. Harshavardhan, and S. M. Anlage, Appl. Phys. Lett. {\bf 81}, 4979 (2002).
\bibitem{ZhuravelJSNM} A. P. Zhuravel, S. M. Anlage, A. V. Ustinov, J. Supercond. Nov. Magn. {\bf 19}, 625 (2006).
\bibitem{Oates} D. E. Oates, A. C. Anderson, P. M. Mankiewich, J. Supercond. {\bf 3}, 251 (1990).
\bibitem{Yu} A. Yu. Rusanov, M. B. S. Hesselberth, and J. Aarts, Phys. Rev. B {\bf 70}, 024510 (2004).

\end{thebibliography}

\end{document}